ORIGINAL REPORT

# Vacancy-concentration-dependent thermal stability of fcc-(Ti,Al)$N_x$ predicted via chemical-environment-sensitive diffusion activation energies

Ganesh Kumar Nayak[a], David Holec[b], and Jochen M. Schneider[a]

[a]Materials Chemistry, RWTH Aachen University, Kopernikusstr. 10, D-52074 Aachen, Germany;
[b]Department of Materials Science, Montanuniversität Leoben, Franz-Josef-Strasse 18, A-8700, Leoben, Austria

**ABSTRACT**

Thermal decomposition of metastable fcc-(Ti,Al)Nx limits the lifetime of coated components. While energetic decomposition aspects can be modelled reliably, the inherent variability of chemical environment-dependent diffusion activation energies remains systematically unexplored. Here, we predict an activation energy range (envelope) for mass transport in varying chemical environments, reflecting the vacancy concentration range fcc-$(Ti_{0.5}Al_{0.5})_{1-x}N_x$ with x = 0.47, 0.5, 0.53. The stoichiometric compound shows maximum thermal stability, consistent with experimental data. Metal vacancies decrease the average migration energy, while metal and nitrogen vacancies reduce barriers via lattice strain relaxation, enhancing mobility. The strong chemical environment dependence challenges conclusions from single-point activation energy data.

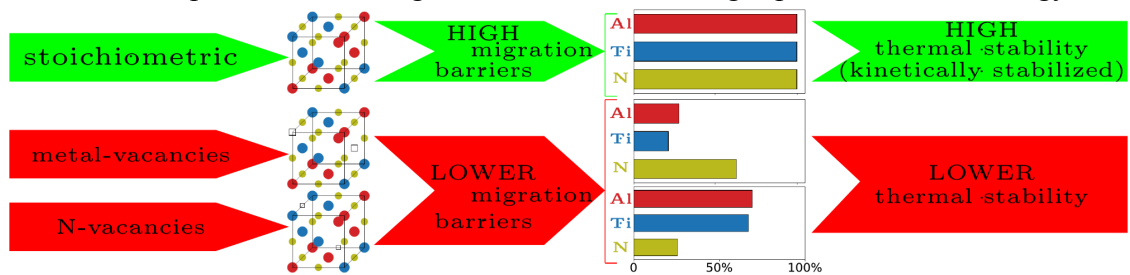

The service life of fcc-(Ti,Al)$N_x$ industrial protective hard coatings is limited by thermally induced decomposition of the metastable solid solution into fcc and hcp phases [1] via spinodal decomposition [2–5]. At around 1000°C, fcc-A1-Al rich domains, formed during spinodal decomposition, transform into hcp, resulting in a concurrent volume increase that can lead to cracking [1, 4]. To enhance the thermal stability of fcc-(Ti,Al)N$_x$, one research direction has focused on point defect engineering [6–8]. However, the impact of point defects on thermal stability remains a topic of debate, with conflicting findings reported.

Alling et al. [7] reported that the mixing enthalpy of stoichiometric Ti$_{1-y}$Al$_y$N$_{1-x}$ is positive but decreases with increasing N vacancy concentration. At high Ti content, introducing N vacancies reduces the mixing enthalpy due to the strong chemical preference of N vacancies to cluster around Ti-rich regions. In contrast, at high Al content, N vacancies increase the mixing enthalpy. Schramm et al. [9] observed that the onset of phase separation is shifted to higher temperatures when reducing the N content in fcc-Ti$_{1-y}$Al$_y$N$_x$ coatings with y = 0.48 − 0.60 and claimed that this trend is consistent with the reduction in mixing enthalpy predicted by Alling et al. [7]. Conversely, to Baben et al. [6] have shown that actively reducing point defects (both metal and N) is critical to achieve high thermal stability. While the energetics of the decomposition process are comparatively well described [1–7], its kinetic aspects have only been explored in a limited number of studies [6, 8].

Grönhagen et al. [8] concluded, based on decomposition kinetics phase-field modeling, that the onset temperature of spinodal decomposition decreases with increasing vacancy concentration. In this study, only metal vacancies were considered, and the diffusion of Al and Ti was modeled during spinodal decomposition only for vacancy concentrations of $10^{-6}$ and $10^{-5}$. They reported that Al diffuses faster than Ti due to the asymmetric curvature of the Gibbs energy curve of vacancies, which is larger on the Al-rich side, even though the diffusivities of Al and Ti were assumed to be equal in the model. The formation of the hcp phase was not considered in this study.

Utilizing a correlative theoretical and experimental research strategy to Baben et al. [6] related the population of point defects to diffusivity and therefore thermal stability. The authors report that the thermal stability of close to stoichiometric (Ti,Al)N thin films was significantly larger compared to over- and understoichiometric variants. Using ab initio calculations to determine point defect energies combined with thermodynamic modeling of defect concentrations under varying reactive gas partial pressures, the enhanced thermal stability of the (Ti,Al)N variant with minimized point defects was explained: This variant exhibited a diffusivity reduced by 12 orders of magnitude compared to the variant with the highest defect concentration [6]. The kinetic aspects of this study were based on calculating the diffusion migration energy for one selected migration pathway only (single-point-calculated), despite the presence of numerous variations in the chemical environment. While a single-point-calculated activation energy is relevant for the corresponding chemical environment, it can not capture the complex variations in chemical environments in a metastable solid solution crystal in the presence of point defects.

In this work, we unravel the effect of point defects on the thermal stability of fcc-(Ti,Al)N$_x$, explicitly accounting for the intrinsic variability of chemical environment-sensitive diffusion activation energies for the first time. The activation energies $(E_a)$ are evaluated from the vacancy formation energies $(E_f)$ and the diffusion migration energy barriers $(E_b)$, using a computational approach previously introduced as the "envelope" method: This method, based on 0 K DFT calculations, was previously employed to predict the Al-rich domain in stoichiometric (V,Al)N [10] and to investigate the underlying interplay among local chemistry, kinetics and lattice distortion in stoichiometric, chemically complex nitrides [11].

As revealed by to Baben et al. [12], metal vacancies enable N overstoichiometry, while N vacancies lead to N understoichiometry. It has been shown before that the concentration of vacancies in fcc-B1 random alloys is remarkably higher than that of the other defects, such as interstitials and antisites [13]. Therefore, only both types of vacancies were included in our calculations. We considered four different cases of fcc-$(Ti_{0.5}Al_{0.5})_{1-x}N_x$ with x = 0.47, 0.5, 0.53. The structures are based on a 64-site fcc B1 supercell, comprising separate metal (Ti and Al) and N sublattices, each containing 32 atoms in the defect-free configuration. To obtain the target compositions, the special quasi-random structure (SQS) method [14] was employed to model random solid solutions by distributing metals with and without vacancies, or N with and without vacancies, across their respective sublattices. To this end, vacancies are introduced into the fcc-$(Ti_{0.5}Al_{0.5})_{1-x}N_x$ with x = 0.5 configuration containing 32 metal and 32 N sites to calculate the energy of formation $(E_f)$ of a particular vacancy in a particular chemical environment building on the method employed for stoichiometric configurations from [10, 11]:

$$E_f = E_i - E_0 + \mu_i \qquad (1)$$

where $E_i$ and $E_0$ are the total energy of the supercell with and without the vacancy, respectively, and $\mu_i$ is the chemical potential of the species $i$. The reference energy for each element was conventionally set to the energy per atom in its most stable form: fcc-A1-Al, hcp-Ti, and the N2 molecule. Consequently, 64 individual vacancy calculations were performed for each system. The predicted variation in vacancy formation energies $(E_f)$ across atomic sites underscores their strong sensitivity to the local chemical environment. Subsequently, diffusion migration energy barriers $(E_b)$, dependent on the local chemical environment, were calculated using the nudged elastic band (NEB) method [15]. These calculations reveal the likelihood of atomic migration between neighboring sites, offering insight into the feasibility of diffusion in chemically varied regions within the solid solution. Thereby, the envelope approach [10, 11] ensures that an atom migrates to a vacant site only if that site was originally occupied by the same atomic species in the parent (vacancy-free) supercell. This constraint enhances computational efficiency by allowing all configurations to be systematically derived from a single parent structure.

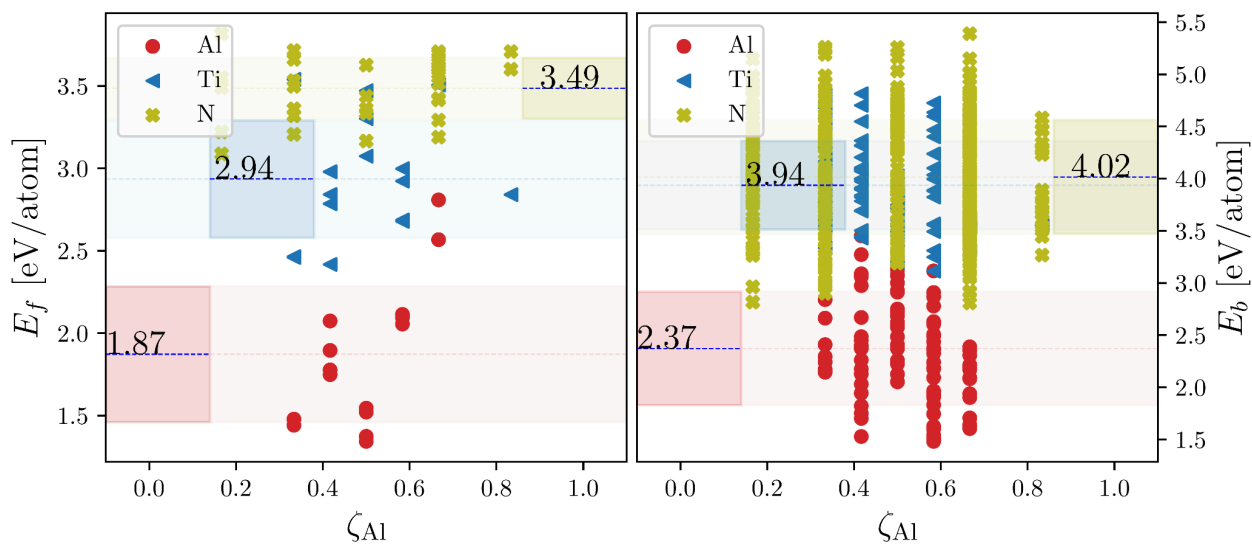

**Figure 1**. Envelopes for (a) vacancy formation energies ($E_f$) and (b) diffusion migration energies ($E_b$) of all species for fcc-$Ti_{0.25}Al_{0.25}N_{0.5}$, plotted as a function of the fraction of Al neighbors, $\zeta_{Al}$, which occupy the 1st nearest neighbor (1NN) sites of N atoms and the 2nd nearest neighbor sites (2NN) for metal atoms.

To compute the formation energy, $E_f$, and migration barrier, $E_b$, we employed density functional theory (DFT) as implemented in the Vienna Ab initio Simulation Package (VASP) [16, 17]. While the computational methodology for the vacancy-free supercell, in terms of chemical environment, follows our previous work [10, 11], the approach for vacancy-containing parent configurations with x = 0.47, 0.53 differs by explicitly accounting for preexisting vacancies. The estimation of $E_f$ was achieved as described above; for instance, the x = 0.53 configuration contains 32 metal sites, 28 of which are occupied by metals and 4 by vacancies, along with 32 N sites. The computation of $E_b$ follows the same constraint as above, except that additional preexisting vacancies are available.

To establish a reference configuration for discussing the vacancy concentration–dependent thermal stability, we begin with the fcc-$Ti_{0.25}Al_{0.25}N_{0.5}$ composition, in which the metal sublattice comprises 16 Ti and 16 Al atoms. The calculated lattice parameter is 4.17 Å, closely matching the previously reported value of 4.19 Å [1]. Since compounds with fully populated metal and nonmetal sublattices lack structural vacancies, mass transport relies on thermally generated vacancies, as described in Eq. (1). The energies required to form a single vacancy in fcc-$Ti_{0.25}Al_{0.25}N_{0.5}$ are collected in Figure 1 (a), where each species is represented with distinct markers and color codes. Horizontal dotted lines indicate the mean Ef for each species, and shaded regions reflect the standard deviation due to changes in the local chemical environment across the solid solution, expressed as the fraction of Al neighbors, $\zeta_{Al}$, which occupy the 1st nearest neighbor (1NN) sites of N atoms and the 2nd nearest neighbor sites (2NN) for the metal atoms. This graphical representation, capturing the impact of chemical variability on vacancy formation energies, is consistently applied to all subsequently discussed compositions.

As shown in Figure 1 (a), in fcc-$Ti_{0.25}Al_{0.25}N_{0.5}$, Al exhibits the lowest $E_f$ among all species, making Al vacancies the most likely to form during heat treatment. Furthermore, the limited overlap between the Al and Ti $E_f$ envelopes indicates that Ti vacancies typically form only after a substantial number of Al vacancies have already developed. N vacancies require even higher formation energies and appear only with further thermal activation. This suggests a vacancy formation sequence of Al → Ti → N.

Figure 1 (b) presents the corresponding $E_b$ envelopes, representing the chemical-environment-dependent energy barriers that atoms must overcome to migrate to adjacent vacant lattice sites. From the comparison of the Eb envelope overlaps between the different migrating species, a similar picture than for the vacancy formation sequence emerges: Al migration is enabled at lower energies than Ti and N. Experimentally, this trend—where Al-rich domains form before Ti-rich regions—has been confirmed via atom probe tomography [6, 18]. A straightforward way to analyze and compare the envelopes is by computing average values from the distributions and to assess the width of the distribution via the standard deviation, as well as the difference between the minimum and maximum values in each envelope. These data are compiled in Table 1 for the stoichiometric composition as well as for the metal vacancy containing and the N vacancy containing compositions, while the corresponding envelope data for fcc-$(Ti_{0.5}Al_{0.5})_{1-x}N_x$ with x = 0.47, 0.53, formatted as in Figure 1, can be found in supplementary information.

| Composition | Energies [eV/atom] | Metric | Al | Ti | N |
|---|---|---|---|---|---|
| Ti$_{0.25}$Al$_{0.25}$N$_{0.5}$ | $E_f$ | Min | 1.34 | 2.42 | 3.08 |
| | | Mean | 1.87 | 2.94 | 3.49 |
| | | Max | 2.81 | 3.54 | 3.82 |
| | | Std | 0.41 | 0.35 | 0.18 |
| | $E_b$ | Min | 1.48 | 3.11 | 2.81 |
| | | Mean | 2.37 | 3.94 | 4.01 |
| | | Max | 3.45 | 4.81 | 5.39 |
| | | Std | 0.54 | 0.42 | 0.54 |
| Ti$_{0.23}$Al$_{0.23}$N$_{0.53}$ | $E_f$ | Min | 3.64 | 6.08 | 1.71 |
| | | Mean | 4.94 | 7.07 | 2.61 |
| | | Max | 5.95 | 8.07 | 3.14 |
| | | Std | 0.64 | 0.53 | 0.35 |
| | $E_b$ | Min | -1.31 | -1.13 | -1.13 |
| | | Mean | 1.18 | 1.46 | 2.11 |
| | | Max | 3.25 | 4.73 | 5.19 |
| | | Std | 1.10 | 1.48 | 1.56 |
| Ti$_{0.26}$Al$_{0.26}$N$_{0.47}$ | $E_f$ | Min | 0.93 | 1.86 | 2.59 |
| | | Mean | 1.46 | 2.48 | 3.37 |
| | | Max | 2.02 | 3.21 | 3.97 |
| | | Std | 0.32 | 0.32 | 0.26 |
| | $E_b$ | Min | -0.70 | -0.62 | -0.68 |
| | | Mean | 1.47 | 2.13 | 1.91 |
| | | Max | 2.94 | 4.82 | 5.79 |
| | | Std | 1.00 | 1.58 | 1.68 |

**Table 1**. The minimum values (Min), average values (Mean), maximum (Max) values, and standard deviation (Std) of vacancy formation energy, $E_f$, diffusion migration energy, $E_b$.

In Figure 2, the predicted -average- migration activation energy ($E_a$) [18] is presented for each species and composition considered, where $\overline{E}_a = \overline{E}_f + \overline{E}_b$, with $\overline{E}_f$ denoting the -average- vacancy formation energy and $\overline{E}_b$ the corresponding -average- migration barrier. Dashed regions mark compositions where migration does not require thermal vacancy generation. In Figure 2, the effect of introducing metal vacancies on the migration energetics is revealed. When metal vacancies are introduced into the crystal structure of the stoichiometric compound, the average activation energy for migration ($\overline{E}_a$) is reduced to the average migration barrier ($\overline{E}_b$) for both Al and Ti. This substantial reduction – by approximately 72% for Al and 78% for Ti – occurs because thermal vacancy generation is no longer a requirement for atomic migration, as pre-existing metal vacancies, accommodating the N overstoichiometry, enable migration. In addition, the magnitude of $\overline{E}_b$ is reduced relative to that in the stoichiometric configuration. This significant decrease – by 50% for Al and 62% for Ti – is likely attributable to the relaxation of local lattice strain surrounding the metal vacancies, thereby accelerating metal migration and hence, decomposition.

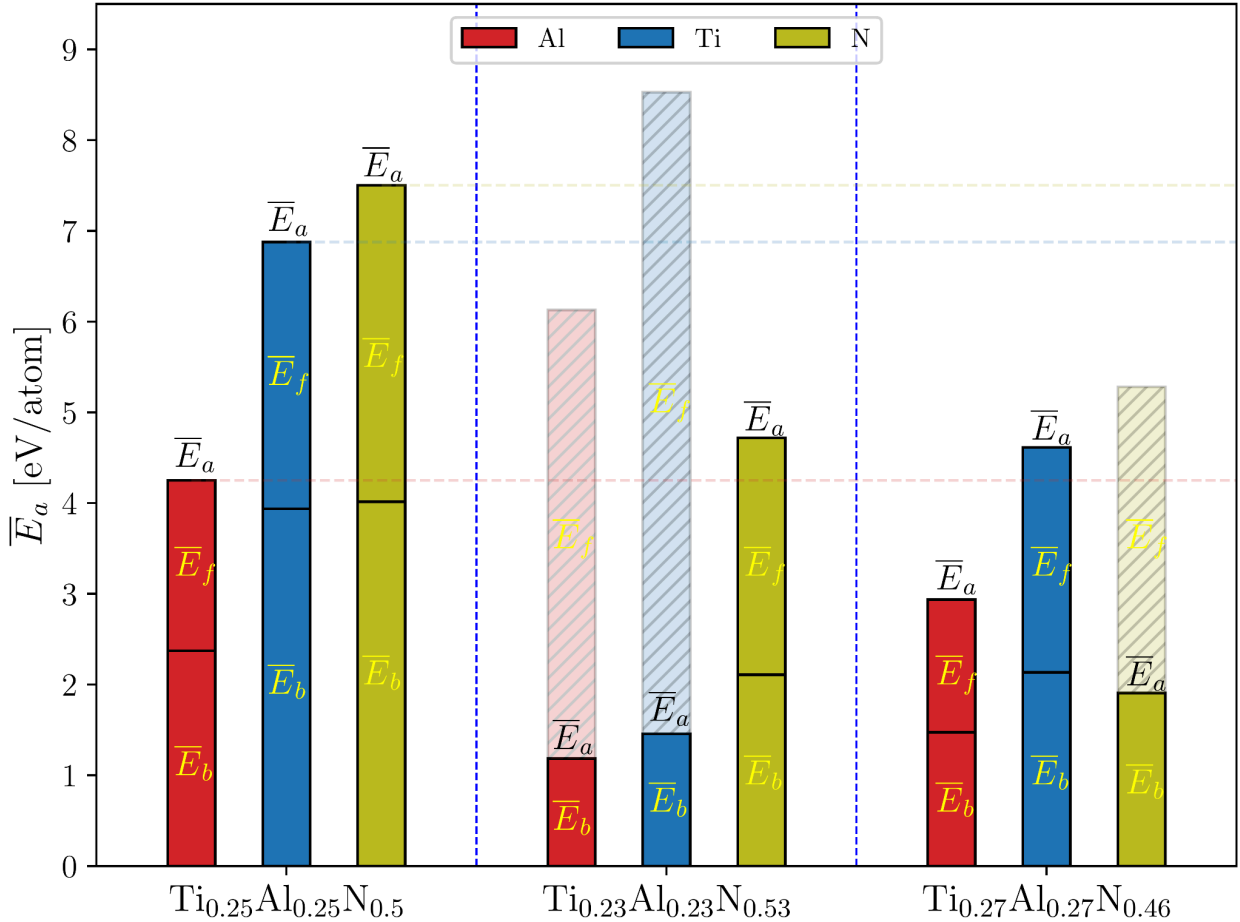

**Figure 2**. The average activation energy for migration $E_a$, is plotted as the sum of the average vacancy formation energy, $E_f$, and the average diffusion migration energy, $E_b$, for all species and all here considered compositions. Dashed regions mark compositions where migration does not require thermal vacancy generation.

Similarly, the introduction of N vacancies leads to a decrease in the average migration barrier for both Al and Ti relative to the stoichiometric variant. This decrease is also likely a result of local lattice strain relaxation near the N vacancies, which facilitates atomic movement by reducing migration resistance, causing a decrease in thermal stability compared to fcc-Ti$_{0.25}$Al$_{0.25}$N$_{0.5}$. An increase in the concentration of metal or N vacancies leads, hence, to a noticeable decrease in thermal stability, as demonstrated by the changes in the maximum diffusion activation energies of Al and Ti within fcc-$Ti_{0.25}Al_{0.25}N_{0.5}$. The predicted reduction in thermal stability–and the associated decrease in decomposition temperature–indicates that metal and N vacancies critically facilitate atomic mobility in the here investigated nitrides.

Consequently, stoichiometric fcc-$Ti_{0.25}Al_{0.25}N_{0.5}$ is predicted to exhibit the highest thermal stability, a conclusion that aligns well with existing experimental data reported in the literature [6]. Fully populated metal and nonmetal sublattices thus appear essential for optimizing the thermal stability of these materials. Analysis of the widths of the predicted energy envelopes reveals a strong dependence of both vacancy formation energies, $E_f$, and migration barriers, $E_b$, on the local chemical environment. This variability–exemplified by the differences between the minimum and maximum values within one envelope (see Figure 1, Table 1, and Figures S1 and S2)–underscores the complexity of 5 atomic migration processes in these materials. It also suggests that activation energy values reported in the literature based on isolated or "single point" calculations may fail to capture the influence of chemical environment heterogeneity on diffusion behavior and, consequently, thermal stability.

In summary, the thermal stability relevant formation energies for vacancies and the corresponding barriers for migration for stoichiometric fcc-$Ti_{0.25}Al_{0.25}N_{0.5}$ have been predicted and compared to variants containing metal vacancies and N vacancies while accounting for the inherent variability of chemical environment. The results can be summarized as follows:

- An increase in metal or N vacancy concentration lowers the decomposition temperature, as indicated by the mean diffusion activation energy of Al and Ti in fcc-$T_{i0.25}Al_{0.25}N_{0.5}$. •
- Thus, the stoichiometric compound is predicted to exhibit maximum thermal stability, consistent with experimental literature data [6].
- The introduction of metal vacancies in the structures reduces the magnitude of the average activation energy for migration, $E_a$, to the average migration barrier, $E_b$, for both Al and Ti, as thermal vacancy generation is not required for migration. Furthermore, the magnitude of $E_b$ is reduced compared to the stoichiometric configuration, likely due to relaxation of local lattice strain in the vicinity of the point defects, thereby accelerating migration.
- From the analysis of the width of each envelope predicted here, it is evident that the magnitude of both $E_f$ and $E_b$ is very strongly chemical environment dependent. Therefore, literature reports based on "single point" activation energy data cannot reflect the complex impact of the chemical environment on the migration behavior.

**Acknowledgement(s)**

The authors highly acknowledge open-access software packages VESTA [19] and pymatgen [20] for supporting the visualization and processing of the structures.

**Disclosure statement**

The authors declare that they have no conflicts of interest or personal relationships that could have appeared to influence the work reported in this paper.

**Funding**

GKN and JMS gratefully acknowledge the Ministry for Culture and Science of the State of North Rhine-Westphalia (MKW NRW) for supporting this work as part of the NHR funding. The authors gratefully acknowledge the computing time provided to them on the high-performance computer CLAIX at the NHR Center NHR4CES with project id p0020883.